# Alternative equation of motion approach applied to transport through a Quantum Dot

**Grzegorz Górski, Jerzy Mizia and Krzysztof Kucab**

Faculty of Mathematics and Natural Sciences, University of Rzeszów

ul. Pigonia 1, 35-959 Rzeszów, Poland

ggorski@univ.rzeszow.pl; mizia@univ.rzeszow.pl

*Abstract* – We study non-equilibrium electron transport through a quantum dot coupled to metallic leads. We use an alternative equation of motion approach in which we calculate the retarded Green function of the impurity by differentiating Green functions over both time variables. Such an approach allows us to obtain the resonance Kondo state in the particle-hole symmetric case and in the asymmetric cases. We apply this technique for calculating the density of the states of quantum dot and the differential conductance as a function of bias voltage. The differential conductance dependence on temperature and on Coulomb interaction is also calculated.

## 1. Introduction

Electronic transport through quantum dots (QD) or single electron transistors (SET) has recently been widely investigated on both the experimental [1-5] and theoretical sides [6-11]. It has a large potential of applications in modern electronics (spintronic) based on nanoscopic structures. Quantum dot systems are excellent systems for studying the Kondo effect. They possess a resonance peak at the Fermi energy in the dot density of states and a zero-bias maximum in differential conductance.

In theoretical analysis the structure of a single quantum dot connected with two leads is well described by the single impurity Anderson model (SIAM) [12]. This model has an exact solution in the case of the noninteracting system ($U = 0$). At finite values of Coulomb interaction $U$ there is no such solution and there are only approximate results. The SIAM model was solved approximately by different numerical techniques (e.g. quantum Monte Carlo [13], numerical renormalization group [13-15]) and analytical techniques (e.g. second-order perturbation theory, modified perturbation theory [9], noncrossing approximation). To solve the SIAM problem one can also use the equation of motion (EOM) technique [6, 7, 10, 11, 16]. This technique relies on finding the retarded Green function of the impurity, and requires the decoupling scheme breaking the infinite set of Green function equations.

One of the broadly used EOM techniques [6, 7, 10, 11, 16] is the decoupling scheme proposed by Lacroix [17]. This scheme was based on the Heisenberg equations of motion with differentiation over one time variable. There was a truncation in the chain of equations at the second order in hybridization term. In the original Lacroix work [17] there was an additional approximation of infinite Coulomb interaction ($U = \infty$). This approach allowed the narrow peak localized on the Fermi level to be obtained, but its width and height were too small resulting in an underestimation of the Kondo temperature [18]. The Lacroix scheme was used for quantum dots with infinite and finite Coulomb interaction. The weak point of this approach, when used at finite $U$, is the wrong result in the particle-hole symmetric case where Kondo peak vanishes. In effect, the unitary limit for the linear conductance is not reached [19]. Current flow in the symmetric case is small which disagrees with the experimental results showing a large conductance [2]. Away from the symmetric case, where the Kondo peak is small, the original EOM method does not fulfill the Fermi liquid relations and in the result linear conductance has very small values. In our report we develop an alternative EOM approach in which we calculate the retarded Green function of the impurity by differentiating Green functions over both time variables. This differs from the commonly used EOM solution by Lacroix [17] where the time derivative was taken only over primary time variable. Our approach allows us to obtain good results in the EOM method not only in the particle-hole symmetric case but also in the asymmetric cases.

The proposed method will be used for systems in the equilibrium state, as well as for systems with the finite bias voltage $V$ applied between the leads coupled to the QD. At the finite bias voltage we calculate the electric current flowing through the dot and the differential conductance $dI/dV$ dependence on temperature for different energy levels of the quantum dot and different Coulomb repulsions. This quantity is observed experimentally. In the particle-hole symmetric case due to the Kondo effect the differential conductance reaches unitary limit [19]. Initial experiments with quantum dots did not support this result. The unitary limit of Kondo effect was demonstrated after using semiconducting quantum dots [1].

The paper is organized as follows: in section 2 we develop our approach to analyze the single impurity Anderson model. Using a modified EOM approach we obtain expressions for the self-energy and Green function in the presence of Coulomb repulsion. In section 3 we present numerical results based on our approach. From the calculated dot Green function we obtain the density of states on the quantum dot and the differential conductance $dI/dV$ as a function of the bias voltage. Dependence of the differential conductance $dI/dV$ on temperature and on Coulomb interaction is also analyzed. It is shown that the increase of temperature reduces the value of the differential conductance $dI/dV$. At zero temperature and in the particle-hole symmetric case we obtain a unitary limit for conductance.

Our results are compared with the experiment and previous calculations. Final conclusions are given in section 4.

## 2. The model

Using the Anderson-type Hamiltonian we analyze the system that is build out of quantum dot connected to two metallic leads. The Hamiltonian of this model has the form

$$H = \sum_{\sigma}\varepsilon_d \hat{n}_{d\sigma} + U\hat{n}_{d\uparrow}\hat{n}_{d\downarrow} + \sum_{\substack{k\sigma \\ \alpha=L,R}} (\varepsilon_{k\alpha} - \mu_\alpha)\hat{n}_{k\alpha\sigma} + \sum_{\substack{k\sigma \\ \alpha=L,R}} \left(V_{k\alpha} d_\sigma^+ c_{k\alpha\sigma} + h.c.\right), \quad (1)$$

where $d_\sigma^+(d_\sigma)$ are the creation (annihilation) operators for the dot electron with spin $\sigma$, $c_{k\alpha\sigma}^+(c_{k\alpha\sigma})$ are the creation (annihilation) operators for the conduction lead electron, $\alpha = L, R$ correspond to the left and right leads, $\varepsilon_{k\alpha}$ is the energy dispersion of $\alpha$ lead, $\mu_\alpha$ is the chemical potential of $\alpha$ lead, $\varepsilon_d$ is the dot energy, $U$ is the on-site Coulomb interaction between electrons on the dot, and $V_{k\alpha}$ is the coupling between the $\alpha$ lead and the dot.

In our analysis we will use the Green functions method and the equation of motion technique. In the case of non-equilibrium situation the retarded, advanced, and the distribution Green functions have to be calculated. The EOM for Green functions was usually obtained by differentiation over primary time ($t$). For the retarded GF we have

$$i\frac{\partial}{\partial t}\langle\langle A(t); B(t')\rangle\rangle^r = \delta(t'-t)\langle[A(t), B(t')]_+\rangle + \langle\langle[A(t), H]_-; B(t')\rangle\rangle^r \quad (2)$$

and for the distribution Green function we have

$$i\frac{\partial}{\partial t}\langle\langle A(t); B(t')\rangle\rangle^< = \langle\langle[A(t), H]_-; B(t')\rangle\rangle^< . \quad (3)$$

After Fourier transform these expressions become the following equations

$$\varepsilon\langle\langle A; B\rangle\rangle_\varepsilon^r = \langle[A, B]_+\rangle + \langle\langle[A, H]_-; B\rangle\rangle_\varepsilon^r . \quad (4)$$

and

$$\varepsilon\langle\langle A; B\rangle\rangle_\varepsilon^< = \langle\langle[A, H]_-; B\rangle\rangle_\varepsilon^< , \quad (5)$$

which are the commonly used equations of motion form (see e.g. [6, 7, 10, 11, 16, 17]). The deficiency of this approach is the lack of the Kondo resonance on the Fermi energy and unfulfilling the unitary limit for conductance [19] in the particle-hole symmetry case. Out of the particle-hole symmetric case we obtain a narrow Kondo resonance peak, whose height and width are small [18] resulting in an underestimation of Kondo temperature. To solve this problem the authors [18, 20] used equations of motion at higher orders for hopping integral. Roermund and co-workers [20] applying the fourth order equation obtained the Kondo resonance no longer vanishing in the particle-hole symmetric case but the unitary limit was still not fulfilled.

In this paper we will use the alternative EOM approach based on the Heisenberg equation which includes differentiating over the second time ($t'$), resulting in EOM of the following form [21]:

$$-\varepsilon \langle\langle A; B \rangle\rangle_\varepsilon^r = -\langle [A,B]_+ \rangle + \langle\langle A;[B,H]_- \rangle\rangle_\varepsilon^r \tag{6}$$

and

$$-\varepsilon \langle\langle A; B \rangle\rangle_\varepsilon^< = \langle\langle A;[B,H]_- \rangle\rangle_\varepsilon^< . \tag{7}$$

Applying equations (4) and (5) to the noninteracting case one can obtain the exact expression for self-energy. After including Coulomb interaction on the dot there is no exact solution for the self-energy and one has to construct an appropriate approximate solution. Frequently used Ng approximation [22] allows us to replace the distribution Green function by the retarded Green function.

Further on we will use the notation for the Green functions: $G_{d\sigma}(\varepsilon) = \langle\langle d_\sigma; d_\sigma^+ \rangle\rangle_\varepsilon^r$. Using in equation (4) the Hamiltonian (1) and following the method of Kuzemsky [21], (Górski and Mizia [23]), we obtain:

$$[g_{d\sigma}(\varepsilon)]^{-1} G_{d\sigma}(\varepsilon) = 1 + U \langle\langle (\hat{n}_{d-\sigma} - n_{d-\sigma}) d_\sigma; d_\sigma^+ \rangle\rangle_\varepsilon^r , \tag{8}$$

where

$$[g_{d\sigma}(\varepsilon)]^{-1} = \varepsilon - \varepsilon_d - U n_{d-\sigma} - i\Delta(\varepsilon) + i\eta , \tag{9}$$

with $\eta$ being a positive infinitesimal real number and $\Delta(\varepsilon)$ the hybridization coupling defined as

$$\Delta(\varepsilon) = \Delta_L(\varepsilon) + \Delta_R(\varepsilon) = \sum_{\substack{k \\ \alpha=L,R}} \frac{V_{k\alpha}^2}{\varepsilon - \varepsilon_{k\alpha}} . \tag{10}$$

In further analysis we will assume that the hybridization coupling is the same for each electrode ($\Delta_L(\varepsilon) = \Delta_R(\varepsilon)$) and that it is energy independent. The asymmetric coupling with electrodes was analyzed by Krawiec and Wysokiński [11]. In addition we will assume that the hybridization coupling is the unit of energy, $\Delta(\varepsilon) = const \equiv 1$.

To solve equation (8) we have to write the EOM for a higher order Green function $\langle\langle (\hat{n}_{d-\sigma} - n_{d-\sigma}) d_\sigma; d_\sigma^+ \rangle\rangle_\varepsilon^r$. For this function we use equation (6) from which we obtain

$$\langle\langle (\hat{n}_{d-\sigma} - n_{d-\sigma}) d_\sigma; d_\sigma^+ \rangle\rangle_\varepsilon^r [g_{d\sigma}(\varepsilon)]^{-1} = U \langle\langle (\hat{n}_{d-\sigma} - n_{d-\sigma}) d_\sigma; (\hat{n}_{d-\sigma} - n_{d-\sigma}) d_\sigma^+ \rangle\rangle_\varepsilon^r . \tag{11}$$

Following the approach based on the irreducible Green function introduced by Kuzemsky [21] we define the scattering operator

$$P_{d\sigma}(\varepsilon) = U^2 \langle\langle (\hat{n}_{d-\sigma} - n_{d-\sigma}) d_\sigma; (\hat{n}_{d-\sigma} - n_{d-\sigma}) d_\sigma^+ \rangle\rangle_\varepsilon^r . \tag{12}$$

The characteristic feature of the propagator $P_{d\sigma}(\varepsilon)$ is that it cannot be reduced to the lower level functions by any decoupling method [21]. By the help of the Dyson equation we obtain function $G_{d\sigma}(\varepsilon)$ in the form:

$$G_{d\sigma}(\varepsilon) = \frac{1}{\varepsilon - \varepsilon_d - i\Delta(\varepsilon) - U n_{d-\sigma} - \Sigma_{d\sigma}^{'}(\varepsilon) + i\eta} , \tag{13}$$

where $\Sigma'_{d\sigma}(\varepsilon)$ is the self-energy which includes the higher order terms and can be expressed by the scattering operator:

$$\Sigma'_{d\sigma}(\varepsilon) = \frac{P_{d\sigma}(\varepsilon)}{1 + g_{d\sigma}(\varepsilon)P_{d\sigma}(\varepsilon)} \ . \tag{14}$$

According to equation (14) in order to calculate the self-energy $\Sigma'_{d\sigma}(\varepsilon)$ we need the propagator $P_{d\sigma}(\varepsilon)$. Using the EOM method we will obtain the irreducible Green function of the higher order, therefore in the calculations we will use the approximation proposed in Górski and Mizia [23]. Applying Wick's theorem and neglecting the spin-flip and superconductivity terms the propagator $P_{d\sigma}(\varepsilon)$ can be written as:

$$P_{d\sigma}(\varepsilon) = \frac{i}{2\pi} \int_{-\infty}^{\infty} dx \frac{P^{>}_{d\sigma}(x) - P^{<}_{d\sigma}(x)}{\varepsilon - x + i\eta} \ , \tag{15}$$

where

$$P^{>,<}_{d\sigma}(x) \approx P^{0\ >,<}_{d\sigma}(x) = U^2 \frac{1}{(2\pi)^2} \iint g^{<,>}_{d-\sigma}(y) g^{>,<}_{d-\sigma}(z) g^{>,<}_{d\sigma}(x - z + y) dy dz \ . \tag{16}$$

Functions $g^{<,>}_{d\sigma}(x)$ in equation (16) are lesser and greater Green's functions. They can be calculated by the EOM method. Again, to avoid obtaining higher order GF we will use the Ng approximation [22], which will give the following expressions:

$$g^{<}_{d\sigma}(x) = -i 2\pi f_{eff}(x) \operatorname{Im} g_{d\sigma}(x) , \tag{17}$$

$$g^{>}_{d\sigma}(x) = i 2\pi (1 - f_{eff}(x)) \operatorname{Im} g_{d\sigma}(x) , \tag{18}$$

where the effective Fermi function is defined as:

$$f_{eff}(x) = \frac{\sum_{\alpha} \Delta_{\alpha}(x) f(x - \mu_{\alpha})}{\Delta(x)} \ . \tag{19}$$

Neglecting the denominator in (14) and approximating the propagators in (15) and (16) by $P_{d\sigma}(\varepsilon) \approx P^0_{d\sigma}(\varepsilon)$ we obtain for the $\Sigma'_{d\sigma}(\varepsilon)$ expression in the second-order perturbation theory $\Sigma'_{d\sigma}(\varepsilon) = P^0_{d\sigma}(\varepsilon)$. This expression gives good results for a rather small $U$ and in the particle-hole symmetric case ($n_d = 1$). The better form of the self-energy working for larger $U$ and the asymmetric case ($n_d \neq 1$) was obtained in [24]:

$$\Sigma'_{d\sigma}(\varepsilon) = \frac{P^0_{d\sigma}(\varepsilon)}{1 + A P^0_{d\sigma}(\varepsilon)} \ . \tag{20}$$

We will use this expression for the self-energy because, as was shown in our previous paper [23], including the full expression given by (14) together with approximation $P_{d\sigma}(\varepsilon) \approx P^0_{d\sigma}(\varepsilon)$ in (15) and (16) does not allow for reaching self-consistent results at intermediate values of Coulomb repulsion.

The parameter $A$ is the average of $g_{d\sigma}(\varepsilon)$ functions at energies $\varepsilon_d$ and $\varepsilon_d + U$ in the atomic limit and it replaces the function $g_{d\sigma}(\varepsilon)$ in equation (14)

$$A = \frac{1}{2}\left[g_{d\sigma}(\varepsilon_d) + g_{d\sigma}(\varepsilon_d + U)\right]\bigg|_{\Delta \to 0}. \tag{21}$$

## 3. Numerical results

Using self-energy given by (20) we can calculate DOS on the QD by the help of equation

$$\rho_{d\sigma}(\varepsilon) = -\frac{1}{\pi}\operatorname{Im}G_{d\sigma}(\varepsilon). \tag{22}$$

In further analysis we will assume that the metallic leads are nonmagnetic which allows us to drop the spin index. In addition we assume that the average chemical potential of leads fulfill the relation $\mu_L + \mu_R = 0$ and that it depends on the bias voltage $\mu_L - \mu_R = eV$. Quantum dot DOS depends on the Coulomb interaction $U$. In figure 1 we present spectral density of quantum dot in function of the Coulomb interaction strength. Calculations were done in the symmetric case when $\varepsilon_d = -U/2$ and $n_{d\uparrow} = n_{d\downarrow} = n_d/2 = 0.5$. The unit of energy is hybridization coupling parameter $\Delta = 1$. With increase of the Coulomb interaction the three centers DOS structure is becoming more pronounced. It is composed of the quasiparticle resonance peak at the Fermi energy (Kondo peak), and two broad satellite sub-bands corresponding to the levels $\varepsilon_d$ and $\varepsilon_d + U$. Increase of $U$ causes shrinking of the width of the resonance peak, but its value at the Fermi energy, $\rho_{d\sigma}(0)$, remains constant.

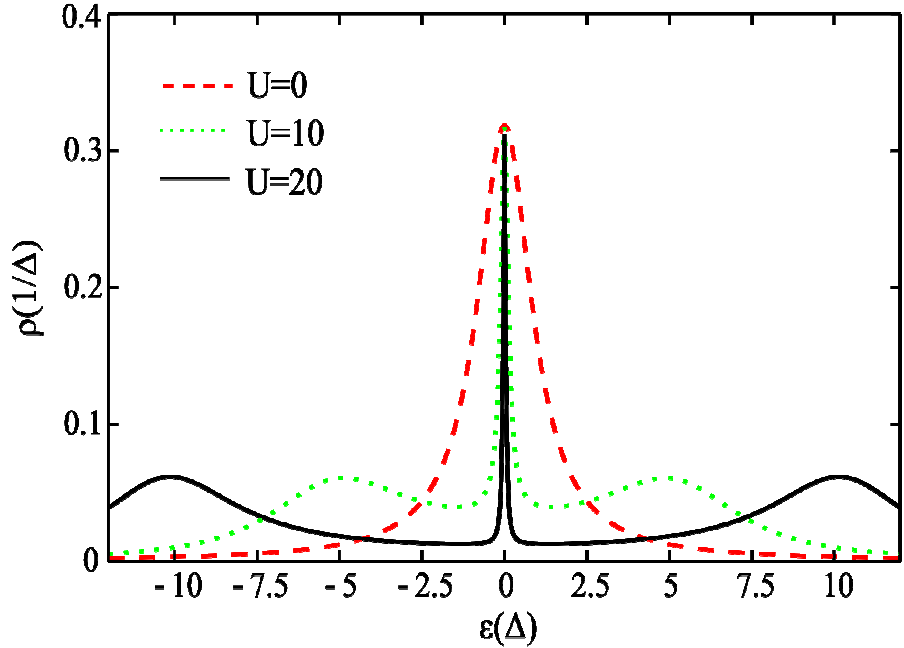

**Figure 1.** Spectral density at the quantum dot as a function of energy $\varepsilon$ calculated in the symmetric case $\varepsilon_d = -U/2$ for different values of Coulomb interaction $U$ at $T = 0$.

Figure 2 shows DOS dependence of the quantum dot on the value of the dot's energy $\varepsilon_d$. Change of $\varepsilon_d$ causes a change in the band position by the hybridization with Hubbard sub-bands. At $\varepsilon_d = -U$ or $\varepsilon_d = 0$ the resonance level merges with the upper or lower Hubbard sub-band, respectively. In these calculations we used $U = 10\Delta$ which allows us to visualize better the three centers density of states. In further calculations we will use smaller Coulomb interaction $U = 5 \div 6\Delta$, which was observed experimentally [4, 25].

The non-equilibrium spectral density for a symmetric junction at $U = 5\Delta$, $T = 0.04\Delta$, and different values of the applied voltage $V$ is plotted in figure 3. The Kondo resonance is destroyed with increase of the applied voltage due to redistribution of spectral weight towards higher energies.

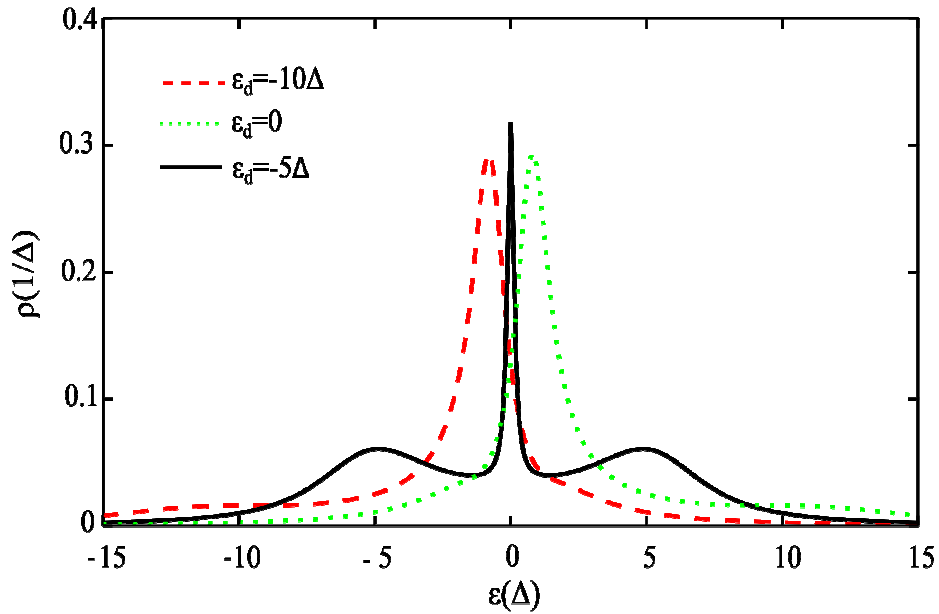

**Figure 2.** Spectral density at the quantum dot as a function of energy $\varepsilon$ calculated for different values $\varepsilon_d$, $U = 10\Delta$ and $T = 0$.

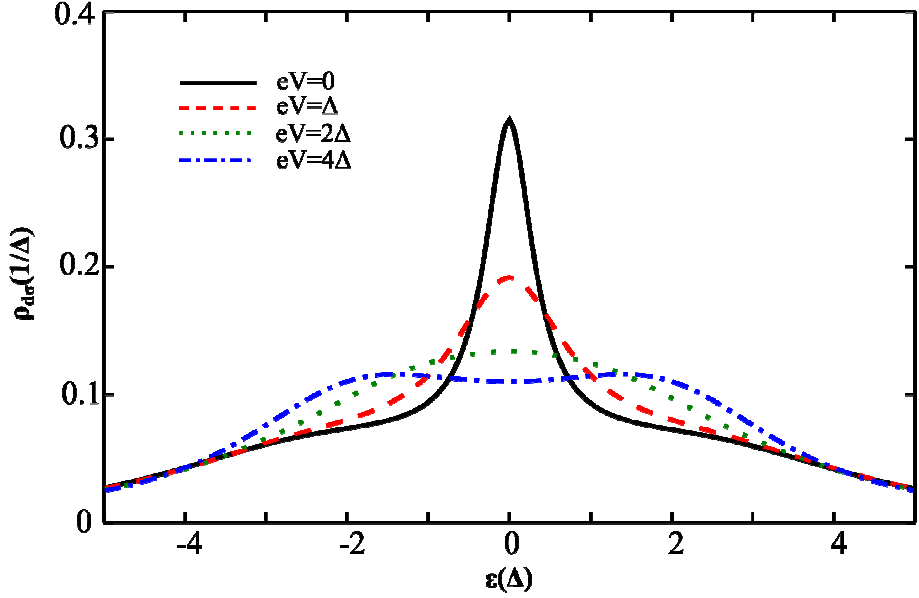

**Figure 3.** Spectral density at the quantum dot as a function of energy $\varepsilon$ calculated in the symmetric case $\varepsilon_d = -U/2$ for different values of the applied voltage between drain and source $eV$, $U = 5\Delta$ and $T = 0.04\Delta$.

The electric current is calculated from the general formula for the current through a region with interacting electrons [26, 27]:

$$I = \frac{4e}{\hbar}\int_{-\infty}^{\infty} \frac{\Delta_L(\varepsilon)\Delta_R(\varepsilon)}{\Delta(\varepsilon)}[f(\varepsilon-\mu_L) - f(\varepsilon-\mu_R)]\rho_{d\sigma}(\varepsilon)d\varepsilon. \qquad (23)$$

Equation (23) allows us to obtain the differential conductance $dI/dV$ by numerical differentiation of the total current. Figure 4 shows the temperature dependence of the differential conductance $dI/dV$ as a function of the applied voltage between drain and source in the symmetric case ($\varepsilon_d = -U/2$). There are three visible maxima, the strongest one is responsible for the zero bias voltage and it is the linear conductance. The remaining two satellite maxima are in the vicinity of $eV \approx \pm U$. Increase of temperature causes a strong reduction of the central peak. This reduction is smaller for the satellites. A strong reduction with temperature of $dI/dV$ at $eV = 0$ is caused by the fact that $dI/dV$ depends on the DOS of the central peak at Fermi level which decreases fast with temperature. At $eV \approx \pm U$ the value of differential conductance $dI/dV$ depends on the DOS of the Hubbard sub-bands which are only weakly changing with temperature. A strong reduction of conductance in the central peak with increasing temperature is observed in experiments ( see e.g. [1, 2, 5]). The weaker reduction for satellite peaks $dI/dV$ is also confirmed by the experimental data [1].

Figure 5 shows the dependence of $dI/dV$ as a function of the applied voltage for different values of Coulomb interaction in the symmetric case. One can see that the maxima of $dI/dV$ are shifted with increasing Coulomb repulsion. It can be also noticed that the central peak narrows when the Coulomb interaction grows. A similar dependence as $dI/dV$ on applied voltage has the density of states of quantum dot on energy (see figure 1). Our results are comparable with those of Han and Heary [13] calculated using the QMC algorithm. The experimental results obtained by Kretinin et. al. [2] have a similar character. They have shown that the increase of $U/\Delta$ decreases width of $dI/dV$ dependence on the applied voltage.

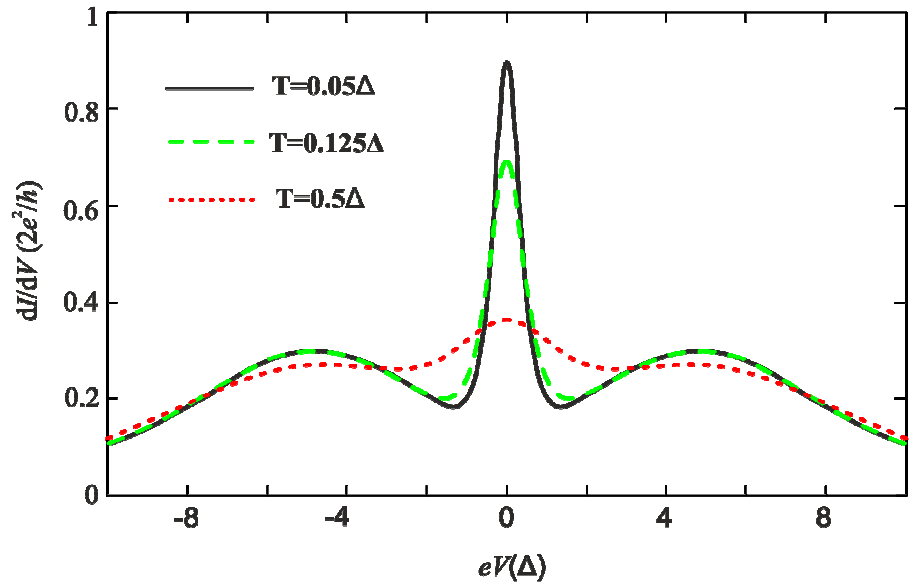

**Figure 4.** Differential conductance $dI/dV$ as a function of the applied voltage between drain and source for different values of temperature ($T = 0.05\Delta$ - solid black line, $T = 0.125\Delta$ - dashed green line, $T = 0.5\Delta$ - dotted red line) and at Coulomb interaction $U = 6\Delta$.

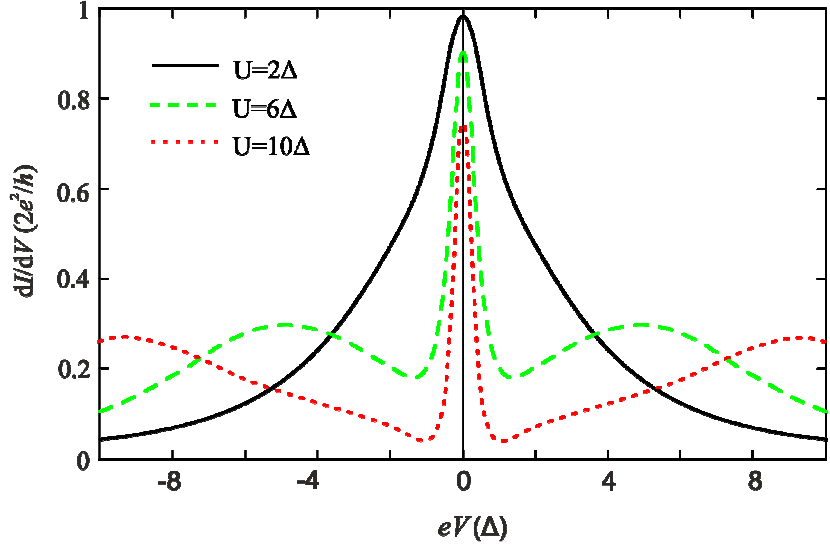

**Figure 5.** Differential conductance $dI/dV$ as a function of the applied voltage between drain and source for different values of Coulomb interaction at temperature $T = 0.05\Delta$ ($U = 2\Delta$ - solid black line, $U = 6\Delta$ - dashed green line, $U = 10\Delta$ - dotted red line).

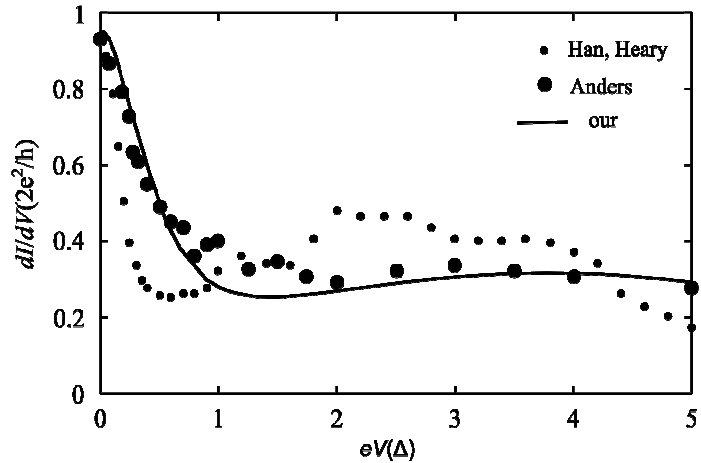

**Figure 6.** Differential conductance $dI/dV$ as a function of the applied voltage between drain and source for Coulomb interaction $U = 5\Delta$ and temperature $T = 0.04\Delta$. The figure shows the comparison between our results and the S-NRG calculations (Anders [15]) and QMC calculations (Han and Heary [13]).

In figure 6 we compare our results for differential conductance at $U = 5\Delta$ and temperature $T = 0.04\Delta$ with the results from the steady-state numerical renormalization group (S-NRG) method [15] and Quantum Monte Carlo (QMC) method with the use of analytic continuation to imaginary Matsubara voltages [13]. Our results agree quite well with the S-NRG calculations [15], especially at small values of the applied voltage and at values close to Coulomb interaction (near Coulomb blockade point). Results obtained by the QMC calculations [13] are not as close to our results or to the results of numerical renormalization group calculations. The corrections to the Quantum Monte Carlo introduced in the imaginary-time Quantum Monte Carlo technique by Han [28] have shown better agreement with the S-NRG calculations [15] and with our results.

In figure 7 we show the dependence of differential conductance at zero bias voltage $G_0 = dI/dV (V = 0)$ as a function of the quantum dot energy $\varepsilon_d$ for different values of Coulomb interaction $U$. In the symmetric case $\varepsilon_d = -U/2$ the value of the differential conductance is $G_0 = 2e^2/h$ and it does not depend on Coulomb interaction. Such a result agrees with the unitary limit ($G_0 = 2e^2/h$ for $T = 0$) visible in experiment [1]. In the case of electron-hole asymmetry the conductance is reduced. When the discrete energy level of quantum dot is at $\varepsilon_d$ or $\varepsilon_d + U$ and it aligns with the Fermi level, the two additional peaks appear. Reduction of the conductance by Coulomb interaction is weaker at these energies.

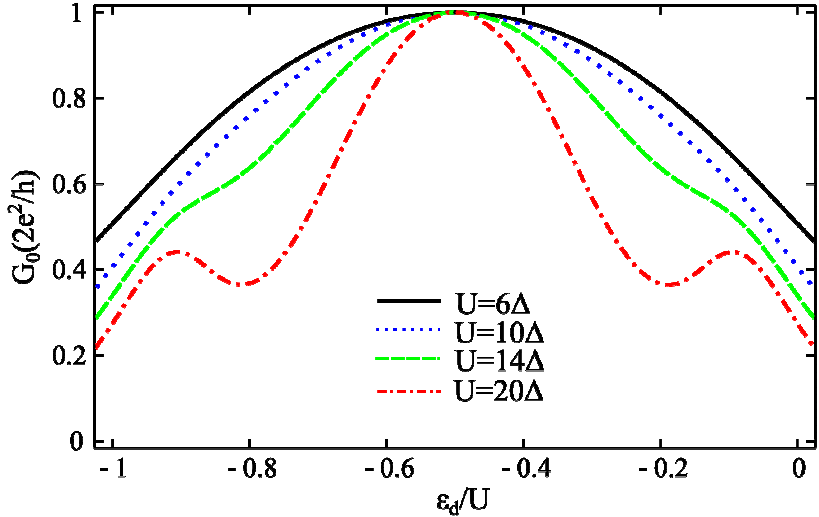

**Figure 7.** Zero-bias differential conductance $G_0 = dI/dV (V = 0)$ as a function of the QD energy $\varepsilon_d$ for different values of Coulomb interaction $U$ and $T = 0$.

The reduction in $G_0$ and appearance of two additional peaks when the discrete energy level of quantum dot $\varepsilon_d$ or $\varepsilon_d + U$ aligns with the Fermi level can be observed in the temperature dependence of $G_0(T)$ (see figure 8). Similar behavior was calculated theoretically by other groups [8] and was observed experimentally [1,3]. Taking into account the data presented in figure 8 and defining Kondo temperature $T_K$ as fulfilling the condition, $G_0(T_K) = G_0(T = 0)/2$, we can see that at $\varepsilon_d$ close to the particle-hole symmetric case ($\varepsilon_d = -U/2$) the value of $T_K$ is smallest and it grows as it departs from this point. In figure 9 we present the Kondo temperature calculated for different values of the dot energy $\varepsilon_d$, with the linear conductance assumed as above. We also show the quadratic function fitted to the results. This function has the dependence: $\ln T_K \sim \varepsilon_d (\varepsilon_d + U)$, the same as the result of the Kondo temperature obtained by Haldane [29]. A similar dependence of Kondo temperature versus $\varepsilon_d$ was shown experimentally by Kretinin and co-workers [2].

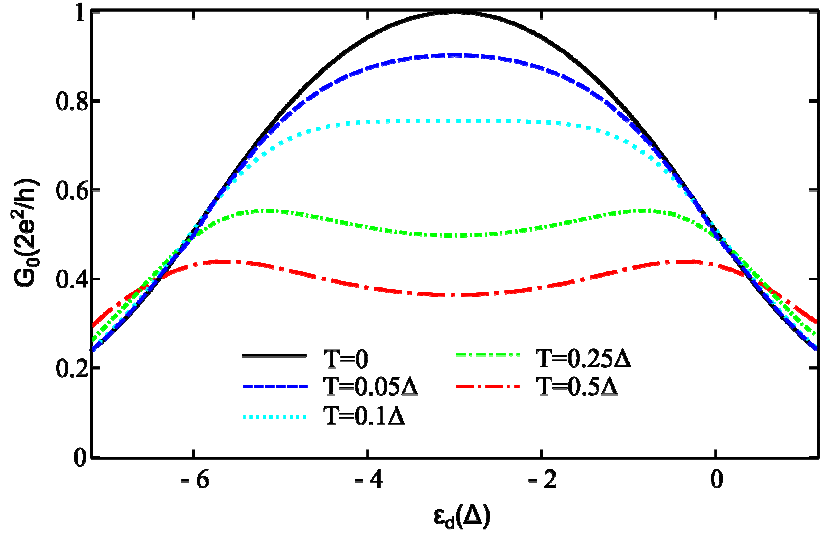

**Figure 8.** Zero-bias differential conductance $G_0 = dI/dV\,(V=0)$ as a function of energy $\varepsilon_d$ for different values of temperature $T$ and Coulomb interaction $U = 6\Delta$.

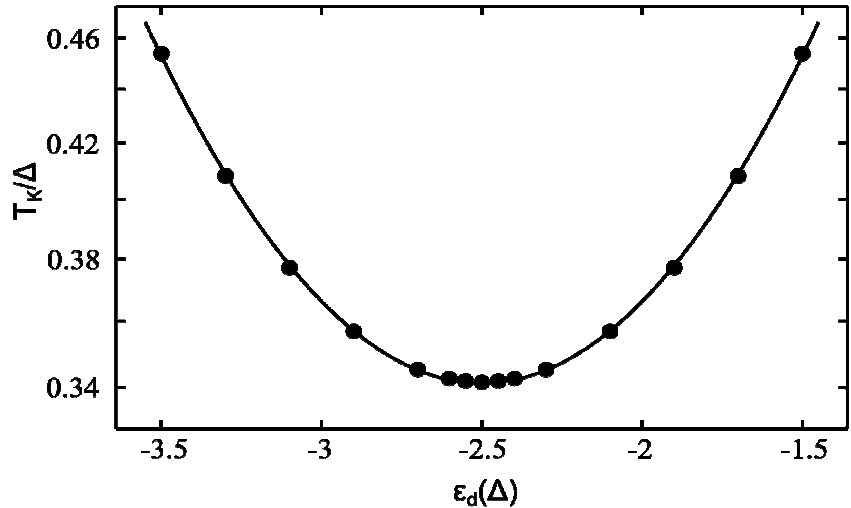

**Figure 9.** Kondo temperature (in logarithmic scale) as a function of the energy $\varepsilon_d$ (dots) for Coulomb interaction $U = 5\Delta$. Solid line presents quadratic function fitted to the results.

In figure 10 we show our results of the linear conductance $G_0 = dI/dV$ ($V = 0$) in function of the $\varepsilon_d$ together with the results from the functional renormalization group (fRG) (large dots) and NRG (small dots) after Jakobs et. al. [30]. Our results are close to the results of the renormalization group (RG) particularly near the Kondo regime (for $\varepsilon_d \approx -U/2 = -3\Delta$) and at the Coulomb blockade regime (when $\varepsilon_d > 0$). At intermediate values of $\varepsilon_d$ the differences between our results and those of the RG method are observed. Comparison of our results with the second-order perturbation theory (2PT) obtained by Jakobs et. al. [30] (not presented in figure 10), shows much better agreement with the RG results, than with the 2PT results.

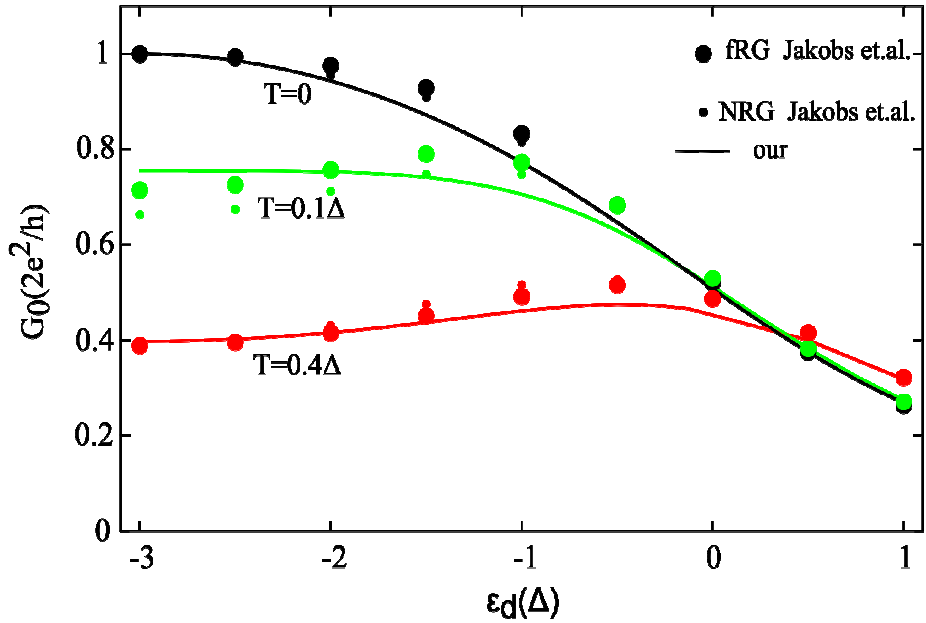

**Figure 10.** Zero-bias differential conductance $G_0 = dI/dV$ ($V = 0$) as a function of the $\varepsilon_d$ for different values of temperature $T$ and Coulomb interaction $U = 6\Delta$. We compare our results with the functional renormalization group (fRG) (large dots) and NRG (small dots) after Jakobs et. al. [30].

The Kondo temperature can be calculated by several methods. As mentioned above we use a method based on the condition for the linear conductance: $G_0(T_K) = G_0(T = 0)/2$. Another frequently used method is estimating the Kondo temperature from half-width at half maximum (HWHM) of the Kondo resonance at T = 0 [31].

In figure 11 we show dependence of the Kondo temperature, $T_K$, on the Coulomb interaction $U$ calculated in the symmetric case $\varepsilon_d = -U/2$ by both methods. Large dots are obtained from relations $G_0(T_K) = G_0(T = 0)/2$, small dots from the half–width at half–maximum (HWHM). At small values of the Coulomb repulsion Kondo temperatures obtained by both methods are close. With a growing $U$ there is an increasing difference between the results. A similar effect was reported by Tosi et. al. [32] who used non-crossing approximation.

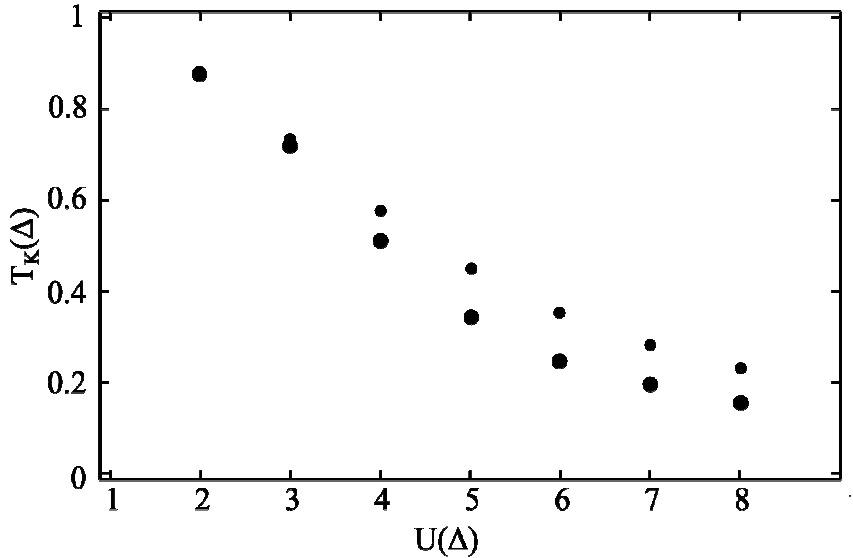

**Figure 11.** Kondo temperature as function of Coulomb interaction in the symmetric case $\varepsilon_d = -U/2$ calculated from relation $G_0(T_K) = G_0(T=0)/2$ (large dots) and from the half–width at half–maximum (HWHM) method (small dots).

**4. Conclusion**

In summary, using an alternative EOM approach we obtained the DOS with the three peaks structure composed of the quasiparticle resonance peak at the Fermi energy (Kondo peak), and two broad satellite sub-bands corresponding to the levels $\varepsilon_d$ and $\varepsilon_d + U$. At $T=0$ the height of the Kondo peak is constant for different values of Coulomb repulsion. Density of states remain the same as in the noninteracting case $\rho_{d\sigma}(0) = \rho^0_{d\sigma}(0)|_{U=0}$. The width of the Kondo peak decreases with increasing Coulomb repulsion. At increasing temperature the maximum of Kondo peak decreases and finally vanishes.

The three centers DOS structure is easily visible in the symmetric case. When the Fermi level, being changed by voltage applied to the quantum dot, approaches one of the Hubbard energies: $\varepsilon_d$ or $\varepsilon_d + U$, the resonance peak hybridizes with the appropriate Hubbard sub-band and disappears.

Similar to the DOS three centre structure is the structure obtained for the differential conductance $dI/dV$. The results show that the linear conductance, responsible for zero bias voltage, strongly decreases with increasing temperature. A similar temperature effect is observed experimentally [3], [1] and was also obtained by other calculations (e.g. [13]). Temperature influence on the satellite maxima of $dI/dV$ is much smaller than on the Kondo peak. Satellite peaks move strongly with the growing Coulomb correlation. Analyzing conductance in the particle-hole symmetric case and at zero temperature we obtain agreement with the unitary limit Kondo effect [19] regardless of the strength of Coulomb repulsion.

Calculations of the differential conductance as a function of applied voltage between drain and source have shown the three centers structure with pronounced central peak depending on temperature and Coulomb repulsion, and with the two satellite maxima. These maxima depend only weakly on temperature but their centres shift strongly apart with growing Coulomb repulsion ($eV = \pm U$). Results agree with the experimental data [1, 2, 5], and they agree very well with numerical calculations by the renormalization group [15] and Quantum Monte Carlo group [13, 28]. Our results of linear conductance in the function of quantum dot energy also show good agreement with those obtained by the renormalization group [30], (see figure 10).

The results obtained here are comparable with the modified perturbation theory (MPT) [9]. What is to our advantage is that they are based on the self-energy $\Sigma'_{d\sigma}(\varepsilon)$ (see equation (20)) which comes directly from the equation of motion for Green functions and not from the interpolation scheme as in the MPT method. In comparison to the old EOM approach [17] we obtain a broad resonance Kondo peak not only in the particle-hole symmetric case but also in the asymmetric cases. Our method of solving the SIAM problem can be applied to the lattice problem within the scheme of the dynamical mean-field theory.